# Photosystem 0, a proposed ancestral photosystem without reducing power that synthesized ATP during light-dark cycling

Anthonie W. J. Muller
*Department of Geology, Washington StateUniversity, Pullman WA 99164-2812*
awjmuller@wsu.com, tel. 509-335-1501

**Photosystem 0 concerns a primitive mechanism for free energy gain as ATP from fluctuating light during early evolution. The PS0 reaction centers had no reducing power: charge transport was only temporary. Light induced within the reaction centers metastable dipoles that generated a membrane potential. This in turn drove ATP synthesis by protons moving through the ATP synthase enzyme. After the decay of the dipole potential in the dark, the protons either (1) returned across the membrane by conduction or (2) were pumped back by ATP synthase, backwards active as ATPase at a higher $H^+$/ATP ratio. PS0 constitutes a link to previously proposed free energy sources for early evolution that worked on thermal cycling. Several contemporary photosynthetic phenomena may be relics of PS0.**

abbreviations: BPS—bacterial photosynthesis; CSP— Charge-Separation Potential: membrane potential due to charge transport across the membrane; LHC—light harvesting complex; MTS—membrane-associated thermosynthesis; PDP—photosynthetic dipole potential: membrane potential due to dipole formation within the membrane; PTS—protein-associated thermosynthesis; PS0— photosystem 0; RC—reaction center; RCII —reaction center of Photosystem II;

## Introduction

A scenario has been proposed for the emergence of bacterial photosynthesis (BPS) in which progenitors of the photosynthetic machinery worked as heat engines during thermal cycling (Muller 1993, 1995, 1996, 1998). Fig. 1 shows the proposed stepwise evolutionary path that started with the $F_1$ moiety of ATP synthase (or, more precisely, its β subunit) and ended with a membrane-diffusible lipid-bound quinone (acquisition of the $bc_1$ complex and light harvesting complex (LHC) is not shown). The heat engines were linked to BPS by Photosystem 0 (PS0), a mechanism that, in the absence of charge transfer across the membrane, worked on light-dark cycling (Fig. 2). The PS0 reaction centers (RCs) formed metastable dipoles in the light that decayed in the dark, and generated a dipole potential that resulted in a membrane potential that in turn drove ATP synthesis in the standard chemiosmotic manner. The PS0 RC evolved from the MTS (membrane-associated thermosynthesis) machinery by stepwise addition of temporary charge carriers (Fig. 3) until it spanned the membrane, or even beyond: the stalk present in many bacterial RCs (BRCs) had in PS0 the function of increasing the RC dipole moment.

The present study elaborates on PS0 and points to its possible relation to many photosynthetic phenomena: metastable RC states, inactive RCs, effects of intermittent light, chloroplast oscillations and State 1-State 2 transitions. A more extensive document on PS0 can be found on the Internet (www.geocities.com/awjmuller/pdf_files/ps0art.pdf).

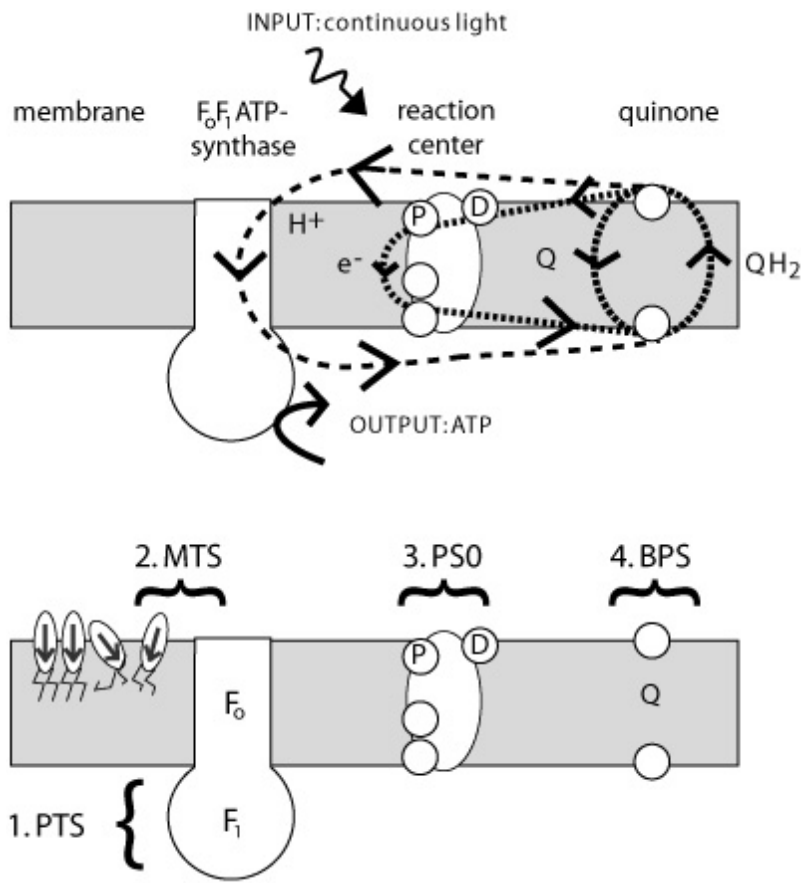

Fig. 1. *Proposed biogenesis of Bacterial Photosynthesis.*
The components of the machinery of bacterial photosynthesis were acquired in the sequence:
1. the $F_1$ part of ATP synthase, during the emergence of protein-associated thermosynthesis (PTS);
2. the $F_o$ part of ATP synthase and the lipids of an asymmetric membrane, during the emergence of membrane-associated thermosynthesis (MTS);
3. the reaction center acquired the stepping stones for the excited electron one after the other during the emergence of photosystem 0 (PS0);
4. membrane-diffusible lipid-bound quinones, during the emergence of bacterial photosynthesis (BPS).

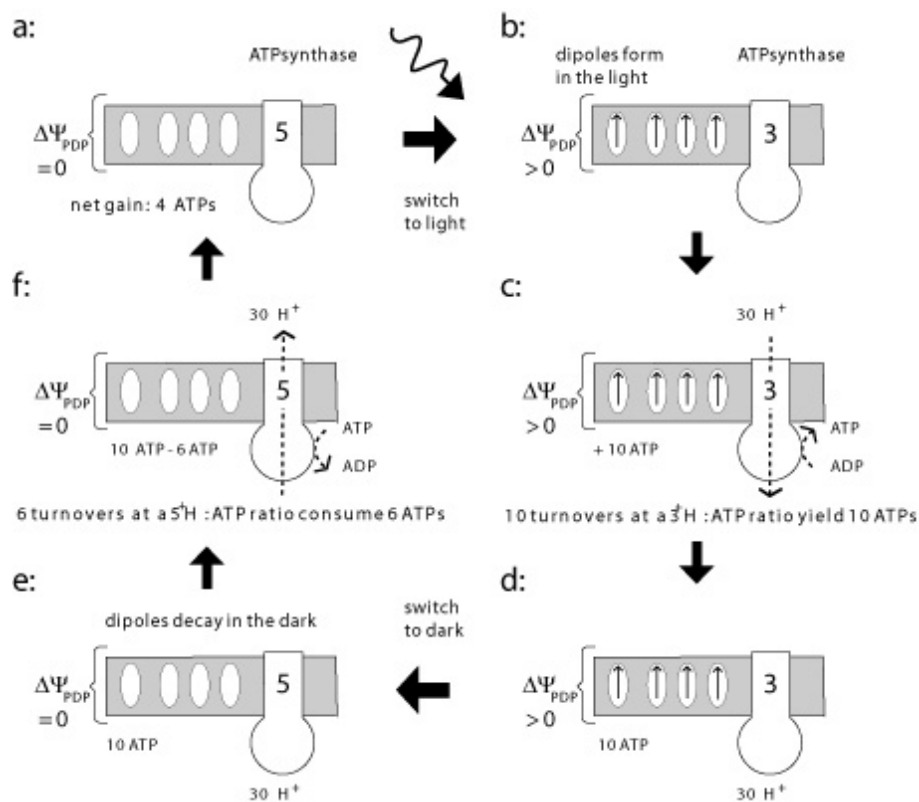

Fig. 2. *Basic principle of the proposed PS0 mechanism.* A membrane contained reaction centers and an ATP synthase that charged and discharged the membrane at different $H^+/ATP$ stoichiometries or modes, here 3 and 5 $H^+/ATP$. Upon illumination (a > b) electric dipoles were formed in the PS0 reaction centers. The dipoles resulted in a dipole potential of the membrane (c), which in turn increased the potential across the membrane. The ATP synthase discharged the membrane in 10 turnovers, transporting 30 protons, yielding, in mode 3, 10 ATP molecules (d). After the light-dark switch the electric dipoles decayed again (d > e), causing the dipole potential to vanish and the membrane potential to drop (e). The ATP synthase charged the membrane at a high $H^+/ATP$ mode of 5. In 6 turnovers, 30 protons were transported, at a cost of 6 ATP molecules (f). At the end of the light-dark cycle (a), a net profit had been made of 10 - 6 = 4 ATP molecules.

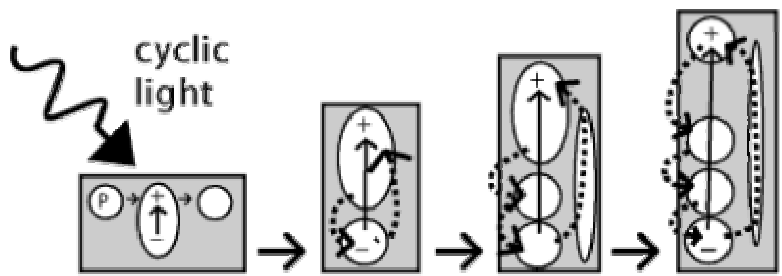

Fig. 3. *The electron transfer chain in the PS0 reaction center evolved by the addition of charge-carrying stepping stones.* The elongated ellipse indicates a distinct return path for the displaced electron that may have evolved simultaneously.

## The Photosystem 0 mechanism

*Fluctuating light* PS0 worked on light-dark cycling, or fluctuating light, which occurs in natural waters due to variable focusing and defocusing of sunlight by surface waves (Schenck 1957; Snyder and Dera 1970). A quickly moving light-dark pattern results that can be seen on the sea bed and at the bottom of swimming pools. The cycle times of these fluctuations range between 0,1 and 10 s.

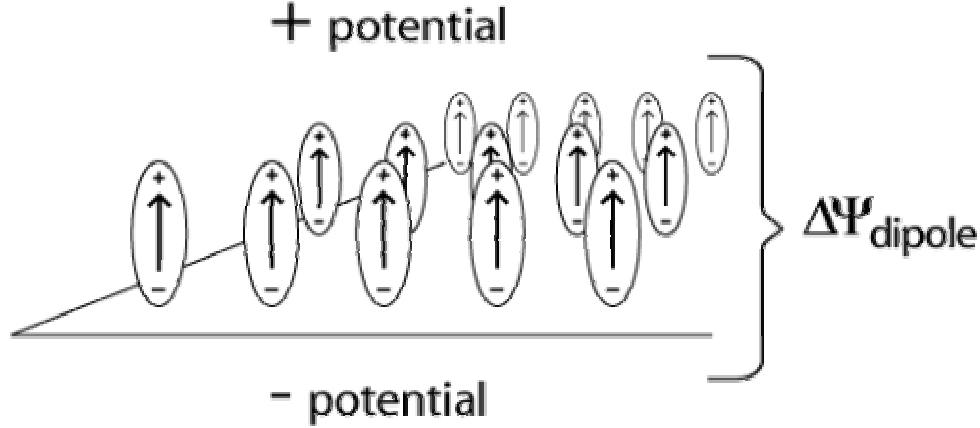

Fig. 4. *The dipole potential*: across a layer of dipoles a potential difference is present.

*The dipole potential* In standard photosynthesis the membrane acts as a capacitor during the generation of the membrane potential $\Delta\Psi$. After a saturating flash, $\Delta\Psi$ equals $\Delta Q / C_m$, with $\Delta Q$ the charge transported across the membrane per unit area, and $C_m$ the electric capacity per unit area (typically ~1 µF/cm$^2$). This $\Delta\Psi$ is called 'Charge-Separation Potential' (CSP), symbol $\Delta\Psi_{CS}$. $\Delta Q$ equals the product of N, the surface density of the RCs, $n_e$, the number of electrons transported per RC per flash, and $e$, the elementary charge; $C_m$ equals $\varepsilon / d$, where $\varepsilon = \varepsilon_r \varepsilon_o$ is the dielectric constant of the membrane and $d$ its thickness. Thus, $\Delta\Psi_{CS} = N\, n_e\, e\, d / \varepsilon_o\, \varepsilon_r$
N is the inverse of A, the membrane area per RC. Hence,

$$\Delta\Psi_{CS} = n_e\, e\, d / A\, \varepsilon_o\, \varepsilon_r \qquad \text{(Junge 1977; Witt 1979)}$$

The potential generated across a layer of dipoles is called 'dipole potential', or 'dipole layer potential' (Fig. 4). The photosynthetic dipole potential (PDP), $\Delta\Psi_{PDP}$, is formed when the electrons excited by a flash halt after a displacement $x$ within the RCs, i.e. inside the membrane. The Helmholtz formula (Jones 1975, p. 38; Schuhmann 1990) then applies, the potential across a layer of dipoles with dipole moment $\mu$, placed at a surface density N in a medium with dielectric constant $\varepsilon$, equals $N\, \mu / \varepsilon$. $\mu$ equals $n_e\, e\, x$. Substitution gives,

$$\Delta\Psi_{PDP} = N\, n_e\, e\, x / \varepsilon_o\, \varepsilon_r = n_e\, e\, x / A\, \varepsilon_o\, \varepsilon_r \qquad (1)$$

Note that $\Delta\Psi_{PDP} / \Delta\Psi_{CS} = x / d$ (Jursinic et al. 1978; Höök and Brzezinski 1994; Popovic et al. 1986). For $x = d$, $\Delta\Psi_{PDP} = \Delta\Psi_{CS}$. Although the values can be identical, the potentials are essentially different: upon illumination with multiple saturating flashes the CSP will continue to increase while the PDP remains constant after the first flash.

*The surface area per RC*, A, follows directly from the number of RC particles per square micron visible in the electron microscope (Staehelin 1986). The lower limit of A equals the surface area of the RC, $A_{RC}$. Removing LHCs decreases $A_{RC}$. A may be 25 nm$^2$ for crystallized central RC cores, 120 nm$^2$ for thylakoids with crystallized RCs (Staehelin 1986), and 150 nm$^2$ for BRCs in chromatophores (Golecki and Oelze 1980). Since the RC particle visible under the electron microscope is a dimer (Boekema et al. 1994; Santini et al. 1994) old reported values of RCII densities in chloroplasts and cyanobacteria (Staehelin 1986) can be doubled, this yields for chloroplasts a value of 450 nm$^2$ for A.

*Estimated PDP values* Eqn (1) can be written as[1],

$$\Delta\Psi_{PDP} [mV] = 18070 \, n_e \, x[nm] / \varepsilon_r \, A[nm^2]$$

The range of possible $\Delta\Psi_{PDP}$ values is wide, from < 5 to 6000 mV: substituting the feasible values of $n_e$ 4, *x* 4, $\varepsilon_r$ 2 and A 25, yields 5800 mV, whereas substituting $n_e$ 1, *x* 1, $\varepsilon_r$ 4,5 and A 500 yields 8 mV. The PDP increases with high $n_e$ and *x*, and low $\varepsilon_r$ and A values.

Comparable CSP values of up to a few hundreds of mV have previously been calculated and observed. For chromatophores Jackson and Crofts (1969) reported a CSP of 430 mV directly after a dark-light switch. Packham et al. (1978) assumed values of $n_e$ 1, *x* 3, $\varepsilon_r$ 3,8 and A 311, which gives 46 mV, the observed value in their single saturating flash experiments. For a single RC turnover, Wraight et al. (1978) calculated a CSP of 100-140 mV. For chloroplasts Witt assumed $n_e$ 2, *x* 3, $\varepsilon_r$ 2, A 1000, resulting in a CSP of 55 mV. Zimányi and Garab (1982) assumed for chloroplasts $n_e$ 1, *x* 7, $\varepsilon_r$ 2 and A 970, which yields 65 mV.

The smaller PDP precedes the CSP, and may approach single-flash CSP values. PDP values higher than 50 mV are feasible for chloroplasts. Where crystallized RCs are present in the membrane, or for chromatophores, even higher values of ~ 200 mV seem feasible.

*ATP synthase /ATPase* The protons driven through ATP synthase by the PDP eventually had to return. This can simply be effected by letting the protons return by conduction. A more interesting but more complex way is to pump the protons back by the same ATP synthase, now active as an ATPase. A symmetry break is involved: for a net gain of ATP the backward reaction has to occur at a higher H$^+$/ATP ratio or 'mode'. In Fig. 2, for example, the two modes are 3 and 5. Fig. 5 shows the proposed regulation of ATP synthase: at high $\Delta\Psi$ active in a low mode, at low $\Delta\Psi$ active in a high mode, and at intermediate $\Delta\Psi$ inactive. This type of regulation was previously proposed for membrane-associated thermosynthesis (Muller 1993). Van Walraven et al. (1997) have reported a similar dependence of the H$^+$/ATP ratio on the light intensity and temperature when these conditions remain constant.

---

[1] $18070 = 10^3 * 1,6 \cdot 10^{-19} * 10^{-9} / 8,854 \cdot 10^{-12} / 10^{-18}$

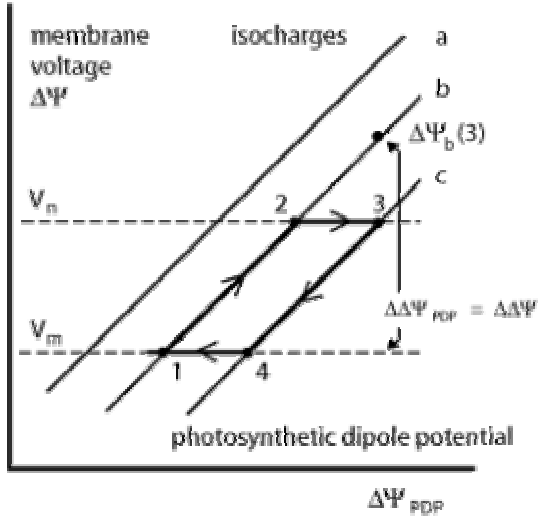

Fig. 5. *The Photosystem 0 cycle*, the membrane voltage - photosynthetic dipole potential ($\Delta\Psi$ - $\Delta\Psi_{PDP}$) cycle of the proposed mechanism. Changes in $\Delta\Psi_{PDP}$ were due to a changing light intensity. The graph of the cycle consists of isocharges and isopotentials. Isocharges (*a*, *b* and *c*) resulted from the linear relation between $\Delta\Psi$ and $\Delta\Psi_{PDP}$ in the absence of charge transfer across the membrane such as ATP synthase activity. The isocharges $V_n$ and $V_m$ depict the equilibrium potentials for mode *n* (*n* $H^+$/ATP) and mode *m* (*m* $H^+$/ATP) activities; $V_n$ and $V_m$ also constitute activity treshold potentials, enzyme activity was switched-on or -off at these potentials. ATP synthase was active either for $\Delta\Psi \geq V_n$ in mode *n*, or for $\Delta\Psi \leq V_m$ in mode *m*; this restricted $\Delta\Psi$ to $V_m \leq \Delta\Psi \leq V_n$.

The cycle started in 1, with $\Delta\Psi = V_m$, where a light intensity increase inactivated mode *m* because $\Delta\Psi$ had increased by the $\Delta\Psi_{PDP}$ increase. Isocharge *b* was followed until $V_n$ was reached (2) and mode *n* was activated. $\Delta\Psi_{PDP}$ continued to rise (total change: $\Delta\Delta\Psi_{PDP}$), but charge transfer by ATP synthase kept $\Delta\Psi$ at $V_n$
(2 > 3). The free energy obtained as ATP by the charge transfer equalled dQ $V_n$, in which
$$dQ = C_m (\Delta\Psi_b(3) - V_n) = C_m \{ \Delta\Delta\Psi_{PDP} - (V_n - V_m) \}$$
was the charge transferred. A subsequent decrease in light intensity in 3 decreased $\Delta\Psi$, which inactivated mode *n*. Isocharge *c* was followed until $\Delta\Psi$ dropped below $V_m$, and mode *m* was activated (4). Along isopotential $V_m$ the membrane was recharged (4 > 1), at a cost of dQ $V_m$, until point 1 was reached again. The net free chemical energy gained as ATP was
$$dQ (V_n - V_m) = C_m \{ \Delta\Delta\Psi_{PDP} - (V_n - V_m) \} (V_n - V_m).$$
As example, consider $\Delta\Delta\Psi_{PDP}$ of 200 mV, with *n* and *m* values of 3 and 5 ($V_3$ = 157 mV, $V_5$ = 94 mV, see text): this yields 9 nJ/cm$^2$ for the work done on the ATP/ADP system in one PS0 cycle.

In practice the regulation of ATP synthase is indeed complex, depending strongly on $\Delta\mu_{H^+}$. At low $\Delta\mu_{H^+}$, ADP binding can inactivate chloroplast ATP synthase, but at a high $\Delta\mu_{H^+}$ the enzyme can be reactivated with simultaneous release of the bound ADP (Junesch and Gräber 1987). Provided the mode changes appropriately, the $\Delta G_P$ and redox state may regulate this enzyme similar to as proposed here (Muller 1993).

*The Photosystem 0 Cycle with a variable ATP synthase stoichiometry*
Figure 5 gives the trajectory of the membrane voltage $\Delta\Psi$ on a $\Delta\Psi$-$\Delta\Psi_{PDP}$ plot during a PS0 cycle. The cycle is similar to the MTS cycle (Muller 1993), the PDP replacing the temperature as variable. Both the $\Delta\Psi_{PDP}$, and the membrane voltage $\Delta\Psi$ rose with the light intensity until ATP synthase was activated in a low mode. Further increase in $\Delta\Psi_{PDP}$ resulted in a discharge of the membrane through the ATP synthase, which stabilized $\Delta\Psi$ at the low mode activation potential. Lowering the light intensity caused $\Delta\Psi_{PDP}$, and as a consequence, $\Delta\Psi$, to decrease, inactivating the low mode. Upon a further $\Delta\Psi$ decrease the enzyme was reactivated in the backwards direction in the high mode, which stabilized the low $\Delta\Psi$ until the end of the cycle.

The cycle time may have been be as small as seconds, since electron transport within the RC occurs in milliseconds, and *in vivo* the ATP synthase of chloroplasts can be activated and deactivated within milliseconds and seconds, respectively (Harris and Crofts 1978; Inoue et al. 1978). At this cycle time, the power of PS0 would be larger than the power of membrane-associated thermosynthesis, where the cycling time is the one due to convection, and estimated to be larger than 10 s (Muller 1993). On the other hand the power of PS0 shall be smaller than the power of BPS, as the comparable cycle time in the latter is the diffusion time of the quinone across the membrane, which is in the millisecond range.

*Evolution of PS0 from MTS*
The proposed MTS worked on thermal cycling, caused by convection in a natural water such as a volcanic hot spring. The temperature changes resulted in thermotropic phase transitions of the membrane, which in turn changed the dipole potentials of the two lipid monolayers that constituted it. Such lipid dipole potential changes of monolayers have been well documented for lipid monolayers at the water/air interface (MacDonald and Simon 1987). In MTS, using an symmetric biomembrane, the resulting membrane potential drove ATP synthesis in the standard chemiosmotic manner, using the same ATP synthase with a variable $H^+$/ATP ratio as proposed here for PS0 (Muller 1993).

MTS is assumed to have yielded the scaffolding for the emergence of PS0 (Fig 1) as follows. The first PS0 system consisted of a membrane protein that contained a single pigment, for instance a chlorophyll molecule. During the MTS convection cycle the pigment was subject to cyclic illumination because of the higher light intensity near the surface of the convection cell. In daylight a single chlorophyll molecule will absorb a photon about every second (Clayton 1980). The excited state resulting from light absorption was stabilized by the formation of a triplet state, which has a lifetime in the range of seconds (Hoff 1986). Excited states having a higher polarizability, the pigment would contribute to the net polarization of the membrane, a contribution varying in synchrony with the convection cycle. A macroscopic energy-conversion device similarly working on polarization changes of a dielectric due to light-induced excited states has previously been proposed (Glazebrook and Thomas, 1982). In this combined MTS/PS0 system the contribution of the excited state of a pigment to the overall dipole potential can at first only have been small. By addition of charge carriers to the mentioned protein, and allowing for transfer of the excited electron to these carriers, larger light-induced dipoles were obtained (Fig. 3). In this way the contribution of light-induced polarization changes steadily increased with respect to the temperature-induced polarization changes. In the end, temperature cycling was not required for energy conversion any more, although some dependency may have remained in other physiological processes.

**Phenomena during contemporary photosynthesis possibly related to PS0**

*Metastable dipole states in RCs* These states are related to delayed fluorescence, luminescence, and phosphorescence. Metastable states with decay times between 1 ms and 10 s have cycle times similar to that of in situ fluctuating light.

*Contrast by the' two-electron gate'* PS0 benefits from 'contrast', a PDP *vs.* light intensity resembling a step-function. Contrast could be obtained by stabilization in the light and destabilization in the dark of the light-induced dipole, such as effected by the 'two-electron gate' (Vermeglio 1977; Wraight 1977, 1982).

*Inactive RCII* In green algae and chloroplasts a significant RCII fraction is 'inactive', the quinone $Q_A$ cannot reduce the quinone $Q_B$ (Chylla et al. 1987; Chylla and Whitmarsh 1989, 1990; Lavergne and Leci 1993). Inactive centers contribute to the membrane potential. Their decay half-time is on the PS0 time scale range, $t_{1/2} \sim 1,7$ s.

*Etiolation of Chloroplasts* Submitting growing chloroplasts to intermittent light (2-118 min LD) stops their development in the so-called etiolated phase, in which the antenna is smaller (Glick and Melis 1988) and the RC density higher (Akoyunoglou 1977). Such etiolated, or intermittent light thylakoids are however completely active.

*Chloroplast Oscillations* Chloroplasts oscillate on the same time scale range as PS0. "Everything oscillates" (Walker 1992): fluorescence, ATP synthesis and hydrolysis, $\Delta pH$, stacking and swelling, oxygen evolution and $CO_2$ assimilation. PS0-like processes may be involved in the oscillations, or their biogenesis.

*State 1-State 2 Transitions* Both temperature (Weis 1985) and light intensity changes (Rouag and Dominy 1994) have been proposed as causes for state transitions.

*Stacking and Swelling of Chloroplasts* Stacking may effectively remove LHCs from the membrane, increasing the RC density, since the stacked area may not function as capacitor, a dielectric that separates lumen and stroma (Williams 1978; Barber 1979). Inactive RCII is found mainly in the stroma exposed region (Guenther and Melis 1990), active RCII in the grana (Staehelin and Arntzen 1979). Upon heat treatment RCII cores move to the stroma (Staehelin 1986; Anderson and Andersson 1988), effectively increasing the RC density in the stroma exposed membrane. Stacking therefore can decrease the capacitor area and increase the RC density, which would benefit PS0 activity by a PDP increase. The small and short charge separation in PS0 may lessen the chance of damage by highly oxidized or reduced species.

**Discussion**

Many photosynthetic phenomena are reminiscent of PS0 and may be relics of a light-cycling requirement in early metabolism. Another possible metabolic relic is the light-dark switch requiring synthesis of a 17,5 kDa translation intermediate of the D1 protein of RCII, which is not formed during continuous darkness or illumination (Inagaki and Satoh 1992). Similarly, a light-dark switch is required by protochlorophyllide synthesis in etiolated leaves (Sironval et al. 1969). Many photoperiodic phenomena are known, with the molecular mechanisms often still unresolved—PS0 may help in disentangling their ontogeny and phylogeny. Bacterial genomes may contain vestiges of genes that coded for proteins that supported functional reconstructions such as PS0.

PS0 is of interest as a possibly verifiable solution to the unsolved problems (Scherer 1983) of the biogenesis of cyclic photosynthetic electron transport, the nature of the first photosynthetic reaction centers (Blankenship 1992) and why things happen in reaction centers as they do (Gunner 1991).

**Conclusion**

Photosystem 0 allows a stepwise model for the emergence and early evolution of photosynthesis, is a possible ancestor of the chloroplast, and gives a new viewpoint on many observations and experimental findings in photosynthesis.